\documentclass[twocolumn,aps,prd,showpacs,10pt,superscriptaddress]{revtex4-1}
\usepackage{amssymb}
\usepackage{amsmath}
\newcommand{\vect}[1]{\boldsymbol{#1}} 
\usepackage{multirow}
\usepackage{graphicx}
\usepackage[usenames,dvipsnames,svgnames]{xcolor}
\usepackage[normalem]{ulem}
\usepackage{hyperref}
\hypersetup{
pdfnewwindow=true,      
colorlinks=true,        
linkcolor=Blue,         
citecolor=Blue,         
filecolor=Blue,         
urlcolor=Blue           
}

\begin{document}

\title{Thermal transport properties of single-layer black phosphorous from extensive molecular dynamics simulations}

\author{Ke Xu}
\affiliation{College of Water Resources and Architectural Engineering, Northwest A\&F University, 712100 Yangling, China}
\author{Zheyong Fan}
\email{brucenju@gmail.com}
\affiliation{School of Mathematics and Physics, Bohai University, Jinzhou, China}
\affiliation{QTF Centre of Excellence, Department of Applied Physics, Aalto University, FI-00076 Aalto, Finland}
\author{Jicheng Zhang}
\affiliation{College of Water Resources and Architectural Engineering, Northwest A\&F University, 712100 Yangling, China}
\author{Ning Wei}
\email{nwei@nwsuaf.edu.cn}
\affiliation{College of Water Resources and Architectural Engineering, Northwest A\&F University, 712100 Yangling, China}
\author{Tapio Ala-Nissila}
\affiliation{QTF Centre of Excellence, Department of Applied Physics, Aalto University, FI-00076 Aalto, Finland}
\affiliation{Center for Interdisciplinary Mathematical Modeling and Department of Mathematical Sciences, Loughborough University, Loughborough, Leicestershire LE11 3TU, UK}

\date{\today}

\begin{abstract}
We compute the anisotropic in-plane thermal conductivity of suspended single-layer black phosphorous (SLBP) using three molecular dynamics (MD) based methods, including the equilibrium MD method, the nonequilibrium MD (NEMD) method, and the homogeneous nonequilibrium MD (HNEMD) method. Two existing parameterizations of the Stillinger-Weber (SW) potential for SLBP are used. Consistent results are obtained for all the three methods and conflicting results from previous MD simulations are critically assessed. Among the three methods, the HNEMD method is the most and the NEMD method the least efficient. The thermal conductivity values from our MD simulations are about an order of magnitude larger than the most recent predictions obtained using the Boltzmann transport equation approach considering long-range interactions in density functional theory calculations, suggesting that the short-range SW potential might be inadequate for describing the phonon anharmonicity in SLBP. 
\end{abstract}

\maketitle

\section{Introduction}

Black phosphorous is a novel layered material which has fascinating electronic properties \cite{li2014nc,liu2014acs,xia2014nc}. It is a semiconductor and its thermal conductivity is thus mainly controlled by phonons. Thermal transport properties in single-layer black phosphorous (SLBP) have been actively investigated theoretically \cite{zhu2014prb,Ong2014Strong, Qin2015Anisotropic, Jain2015Strongly,qin2016prb,Xu2015Direction,Hong2015Thermal,Zhang2016Thermal}, although only the thermal conductivity $\kappa$ of multilayer phosphorene films with thickness down to about $10$ nm have been experimentally measured \cite{Luo2015Anisotropic,Jang2015Anisotropic,Lee2015Anisotropic,Zhu2016Revealing,smith2017am}. The thermal conductivity is found to decrease with decreasing thickness and is about $10$ and $20$  W/mK in the zigzag and the armchair directions, receptively, for the thinnest ($\approx 10$ nm) samples measured \cite{Luo2015Anisotropic}.

Theoretically, the thermal conductivity of SLBP was mainly computed \cite{ zhu2014prb,Ong2014Strong, Jain2015Strongly,Qin2015Anisotropic,qin2016prb} using the Boltzmann transport equation (BTE) approach where phonon-phonon scattering events are described by anharmonic lattice dynamics. All the studies have confirmed the large anisotropy of the thermal transport in SLBP, i.e., the thermal conductivity in the zigzag direction $\kappa^{\rm zig}$ is a few times larger than that in the armchair direction $\kappa^{\rm arm}$, in accordance with the anisotropic crystal structure of SLBP. However, the exact thermal conductivity values depend sensitively on the cutoff distance in the anharmonic lattice dynamics calculations and the exchange-correlation functionals in the density function theory (DFT) calculations; see Ref. \onlinecite{qin2018small} for a review. 

The BTE based method is less suitable for studying systems with large unit cells, in which case molecular dynamics (MD) based methods are generally more useful. Two parameterizations \cite{jiang2015nt,Xu2015Direction} of the Stillinger-Weber (SW) potential \cite{stillinger1985prb} have been developed for SLBP. Using their parameterization  \cite{Xu2015Direction}  and the equilibrium MD (EMD) method based on the Green-Kubo relation \cite{mcquarrie2000book}, Xu \textit{et al.} \cite{Xu2015Direction} obtained $\kappa^{\rm arm} = 33.0$ W/mK and $\kappa^{\rm zig} = 152.7$ W/mK. Using the parameterization by Jiang \cite{jiang2015nt} and the nonequilibrium MD (NEMD) method directly based on Fourier's law, Hong \textit{et al.} \cite{Hong2015Thermal} obtained $\kappa^{\rm arm} = 63.6\pm 3.9$ W/mK and $\kappa^{\rm zig} = 110.7\pm 1.75$ W/mK, while Zhang \textit{et al.} \cite{Zhang2016Thermal} obtained quite different values: $\kappa^{\rm arm} = 9.89$ W/mK and $\kappa^{\rm zig} = 42.55$ W/mK. The discrepancy between Hong \textit{et al.} \cite{Hong2015Thermal} and Zhang \textit{et al.} \cite{Zhang2016Thermal} is puzzling because both have used the NEMD method and the same potential \cite{jiang2015nt}. The results from the two SW parameterizations also differ significantly and the origin for the difference has not been clarified.

To resolve the discrepancies in the previous works, we compute here the in-plane thermal conductivity of SLBP using three different MD based methods: the EMD and NEMD methods mentioned above and a less often used method called the homogeneous nonequilibrium MD (HNEMD) method proposed by Evans \cite{evans1982pla,evans1990book} in terms of two-body potentials, and generalized to general many-body potentials by some of the current authors \cite{fan2018submitted}. We find that all the three methods give consistent results and the predictions by previous MD simulations \cite{Xu2015Direction,Hong2015Thermal,Zhang2016Thermal} are inaccurate due to various reasons.  Our results also suggest that the different results from the two SW potentials are not due to the different MD methods but different parameterizations.  Thermal conductivities calculated using the parameterization by Jiang \cite{jiang2015nt} are closer to those from BTE predictions based on DFT calculations, but they still differ by several times. We also evaluate the relative efficiency of the three MD based methods. Our study demonstrates the importance of properly considering several technical issues in the use of MD based methods for computing thermal conductivity and highlights the importance of the quality of the interatomic potential in predicting the thermal conductivity. 

\section{Models and Methods }

\subsection{Models}

\begin{figure}[htb]
\begin{center}
\includegraphics[width=0.8\columnwidth]{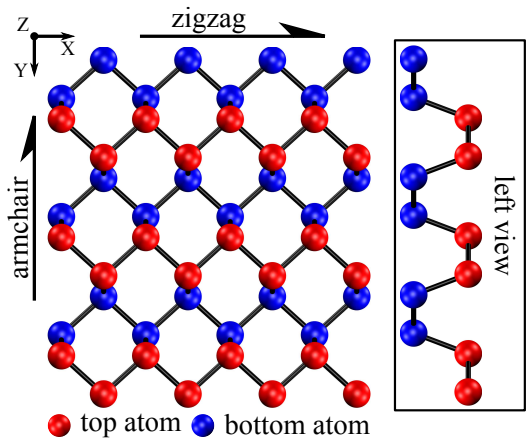}
\caption{A schematic illustration of the atomistic structure of SLBP. Viewed from above, a zigzag edge (horizontal) and an armchair edge (vertical) can be clearly seen. Viewed from left, one can see that the system is puckered along the armchair direction and occupies two layers separated by a given distance. Atoms in the top (bottom) layer are in red (blue).}
\label{f_model}
\end{center}
\end{figure}

A schematic illustration of the atomistic structure of SLBP is shown in Fig. \ref{f_model}. Viewed from the $z$ direction perpendicular to the atomic layer, one can see a zigzag shaped edge along the $x$ direction and an armchair shaped edge along the $y$ direction. Viewed from the side, one can see that the system is puckered along the armchair direction and occupies two layers separated by a given distance. The local environment of an atom from the top layer is different from that of the adjacent atom from the bottom layer. Therefore, when modeling the interactions between the atoms, it is desirable to distinguish between the atoms in the two layers. Jiang \cite{jiang2015nt} and Xu \textit{et al.} \cite{Xu2015Direction} have separately developed a SW potential \cite{stillinger1985prb} in which the atoms from the two layers are treated as different atom types. The SW potential models developed by them are identical except for the different parameterizations. We call the SW potentials parameterized by Jiang \cite{jiang2015nt} and Xu \textit{et al.} \cite{Xu2015Direction} the SW1 and the SW2 potentials, respectively.  

In this work, we only consider isotopically pure and pristine (defect free) SLBP at temperature $T=300$ K and zero pressure. The in-plane lattice constants are determined automatically using a barostat and the thickness of the system is chosen as the conventional value of \cite{qin2016prb} of $0.525$ nm. We only consider heat transport in effectively two-dimensional systems and do not consider edge effects in nanoribbons \cite{liu2017afm}.

\subsection{Methods \label{s_methods}}

We use the open-source GPUMD (Graphics Processing Units Molecular Dynamics) package \cite{fan2013cpc,fan2015prb,fan2017cpc,gpumd} to do the MD simulations. For all the systems, we use the velocity-Verlet integration scheme \cite{tuckerman2010book} with a time step of $2$ fs, which has been tested to be small enough. Because the Debye temperature of LSBP is $278.66$ K according to the calculations in Ref. \onlinecite{Qin2015Anisotropic}, which is smaller than our simulation temperature of $300$ K, there should not be any significant quantum effects on the thermal conductivity of SLBP predicted from classical MD simulations. It is thus justified to use classical MD simulations without explicit quantum corrections.

\subsubsection{The EMD method}

In the EMD method, the running thermal conductivity tensor $\kappa^{\mu\nu}(t)$ is calculated according to the Green-Kubo formula \cite{mcquarrie2000book} as
\begin{equation}
\kappa^{\mu\nu}(t) = \frac{1}{k_{\rm B}T^2V}\int_0^{t} \langle J^{\mu}(0) J^{\nu}(t') \rangle dt',
\end{equation}
where $k_{\rm B}$ is Boltzmann's constant, $T$ and $V$ are respectively the temperature and volume of the system, $J^{\mu}$ is the heat current in the $\mu$ direction, and $\langle J^{\mu} (0)J^{\nu} (t) \rangle$ is the heat current autocorrelation function (HCACF). The HCACF can be calculated from the heat current sampled at equilibrium (hence the name EMD method). For a system of $N$ atoms described by a general many-body potential with the total potential energy
\begin{equation}
U=\sum_{i=1}^N U_i(\{\vect{r}_{ij}\}_{j\neq i}),
\end{equation}
the heat current $\vect{J}$ is (a kinetic term which only matters for fluids is excluded) \cite{fan2015prb}
\begin{equation}
\label{equation:J}
\vect{J} =\sum_i \sum_{j \neq i} \vect{r}_{ij}
\frac{\partial U_j}{\partial \vect{r}_{ji}} \cdot \vect{v}_i,
\end{equation}
where $\vect{r}_{ij} \equiv \vect{r}_{j} - \vect{r}_{i}$ and $U_i$, $\vect{r}_i$, and $\vect{v}_i$ are respectively the potential energy, position, and velocity of atom $i$. 

The EMD method has relatively small finite-size effects and we used a sufficiently large simulation cell consisting of $28000$ atoms, which is about $30$ nm $\times$ $30$ nm in size. Periodic boundary conditions were applied to both the zigzag and the armchair directions. We first equilibrated the system at $300$ K and zero pressure in the NPT ensemble for $2$ ns and then made a production run of $20$ ns in the NVE ensemble. 

\subsubsection{The HNEMD method}

The HNEMD method was first proposed by Evans \cite{evans1982pla,evans1990book} in terms of two-body potentials. Later, Mandadapu \textit{et al.} \cite{mandadapu2009jcp} generalized this method to a special class of many-body potentials (cluster potentials) to which the SW potential belongs. The formalism we present below follows Ref. \onlinecite{fan2018submitted}. In this method, one generates a homogeneous heat current by adding an external driving force (a kinetic term which only matters for fluids is excluded)
\begin{equation}
\vect{F}_{i}^{\rm ext}
= \sum_{j \neq i} \left(\frac{\partial U_j}{\partial \vect{r}_{ji}} \otimes \vect{r}_{ij}\right) \cdot \vect{F}_{\rm e}
\end{equation}
to the interatomic force of atom $i$ resulted from the many-body potential \cite{fan2015prb}
\begin{equation}
\vect{F}_i^{\rm int} = \sum_{j\neq i} \left(\frac{\partial U_i}{\partial \vect{r}_{ij}} - \frac{\partial U_j}{\partial \vect{r}_{ji}} \right)
\end{equation}
to get the total force $\vect{F}_i^{\rm tot}=\vect{F}_i^{\rm ext}+\vect{F}_i^{\rm int}$. The driving force (of dimension inverse length) $\vect{F}_{\rm e}$ should be small enough such that 
the system is in the linear response regime. Quantitatively, it was found \cite{fan2018submitted} that linear response is completely assured when $F_{\rm e} \lambda \lesssim 1/10$, where $\lambda$ can be considered as the average phonon mean free path. Temperature control and momentum conversation need to be taken care of \cite{evans1982pla,mandadapu2009jcp,fan2018submitted}. For temperature control, we use the Nos\'{e}-Hoover chain \cite{tuckerman2010book} method, although the simple velocity rescaling method also suffices. To ensure momentum conservation, one simply needs to correct the external driving force, $\vect{F}_i^{\rm ext} \rightarrow \vect{F}_i^{\rm ext} - (1/N)\sum_i \vect{F}_i^{\rm ext}$, or equivalently, make a similar correction to the total force, because the interatomic forces conserve the total momentum of the system. In this stage, one measures the nonequilibrium heat current $\langle \vect{J} \rangle_{\rm ne}$ where $\vect{J}$ is defined in Eq. (\ref{equation:J}). The thermal conductivity tensor is then calculated according to
\begin{equation}
\frac{\langle J^{\mu}(t)\rangle_{\rm ne}}{TV}
= \sum_{\nu} \kappa^{\mu\nu}(t) F_{\rm e}^{\nu}.
\end{equation}
In practice, one calculates the running average
\begin{equation}
\overline{\kappa}^{\mu\nu}(t)=\frac{1}{t}\int_0^t \kappa^{\mu\nu}(t')dt'.
\end{equation}
and checks its time convergence. More details on this method can be found in Ref. \onlinecite{fan2018submitted}.

The HNEMD method also has relatively small finite-size effects \cite{evans1982pla,fan2018submitted,mandadapu2009jcp,dongre2017msmse} and we used the same simulation cell as in the EMD method. Periodic boundary conditions were again applied to both the zigzag and the armchair directions. We first equilibrated the system at $300$ K and zero pressure in the NPT ensemble for $2$ ns and then switched on the driving force for $10$ ns.  We note that $\kappa^{\rm zig}$ and $\kappa^{\rm arm}$ have to be calculated in different HNEMD simulations with the driving force $\vect{F}_{\rm e}$ applied in different directions, while both of them can be obtained in the same EMD simulation. We chose the magnitude of 
$\vect{F}_{\rm e}$ to be $0.1$ $\mu$m$^{-1}$, which has been tested to be sufficiently small.

\subsubsection{The NEMD method}

\begin{figure}[htb]
\begin{center}
\includegraphics[width=0.8\columnwidth]{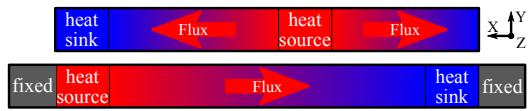}
\caption{Schematic diagram of the periodic boundary setup (upper) and the fixed boundary setup (lower) in the NEMD simulations.}
\label{f_nemd}
\end{center}
\end{figure}

The NEMD method can be used to calculate the thermal conductivity $\kappa(L)$ of a system with a finite length $L$. In this method, a temperature gradient $\nabla T$ is established by generating a nonequilibrium heat flux $Q$ and $\kappa(L)$ is calculated according to Fourier's law as
\begin{equation}
\kappa = \frac{Q} {|\nabla T|}.
\end{equation}
We generate $Q$ by coupling a source region of the system to a thermostat (realized by using the Nos\'e-Hoover chain method \cite{tuckerman2010book}) with a higher temperature of 330 K and a sink region to a thermostat with a lower temperature of 270 K. The heat flux $Q$ can be calculated from the energy transfer rate $d E/d t$ between the source/sink region and the thermostat:
\begin{equation}
Q=\frac{dE/dt}{S},
\label{Q_ex}
\end{equation}
where $S$ is the cross-sectional area perpendicular to the transport direction. Two typical setups in the NEMD method, one with periodic boundaries in the transport direction and one with fixed boundaries, are illustrated in Fig. \ref{f_nemd}. Periodic boundary conditions are applied to the transverse direction in both setups. We note that $S$ has to be taken as twice of the cross-sectional area in the periodic boundary setup. We use both setups and compare them in terms of the results and computational efficiency. 

The length $L$ of the systems considered vary from $200$ nm to $800$ nm with an increment of $100$ nm and the width is fixed to about $20$ nm. As in the HNEMD method, the zigzag and the armchair directions have to be separately considered. For each system, we first equilibrated it at $300$ K in the NVT ensemble for $2$ ns, using the lattice constants determined from the above EMD simulations. Then, we generated the nonequilibrium heat flux using the local thermostats for $20$ ns. We have checked that all the systems have achieved steady state after $15$ ns. Therefore, the temperature gradient and the nonequilibrium heat flux are determined from relevant data within the last $5$ ns in this stage.

\subsubsection{Determination of the uncertainties in the MD results}

The uncertainties in all our simulation results are quantified in terms of the statistical error \cite{haile1992book} from the independent runs. The error is the standard deviation divided by the square root of the number of independent runs. The standard error is the correct indicator of the error bars, which should decrease with increasing number of independent runs. In the EMD method, the signal-to-noise ratio of the HCACF decreases with increasing correlation time \cite{haile1992book} and the resulting integrated thermal conductivity values exhibit large variations from run to run. Therefore, one usually needs to do many independent runs to reduce the uncertainty. In contrast, as one directly measures the heat current in the HNEMD and the NEMD methods, the calculated thermal conductivity values from independent runs show much smaller variations and the number of independent runs needed to achieve an uncertainty comparable to that in the EMD method is much smaller. The number of independent runs used for the EMD, HNEMD, and NEMD methods is respectively 200, 4, and 5. 

\section{Results and Discussion}

\subsection{EMD results}

\begin{figure}[htb]
\begin{center}
\includegraphics[width=\columnwidth]{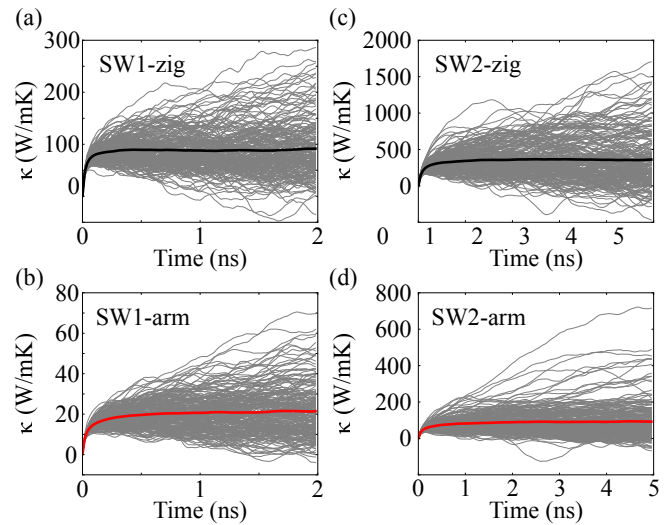}
\caption{Running thermal conductivity of SLBP at $300$ K using the SW1 (a-b) and SW2 (c-d) potentials. The transport directions (zigzag and armchair) are indicated in each subplot. In each subplot, the thin lines represent the results from $200$ independent runs and the thick line represents their average. }
\label{f_emd}
\end{center}
\end{figure}

The running thermal conductivities from the $200$ independent EMD runs (thin lines) and their averages (thick lines) for different potentials (SW1 and SW2) and transport directions (zigzag and armchair) are shown in Fig. \ref{f_emd}. As expected, the variation between the independent runs becomes larger and larger with increasing correlation time because the signal-to-noise ratio in the HCACF becomes smaller and smaller \cite{haile1992book}. The SW2 potential requires a longer correlation time to achieve the convergence of the  running thermal conductivity, which means that the average phonon relaxation time is longer for this potential. For both potentials, we calculate $200$ independent conductivity values at the maximum correlation times shown in Fig. \ref{f_emd} and report their average $\kappa_{\rm ave}$ and standard error (standard deviation divided by the square root of the number of runs) $\kappa_{\rm err}$ in Table \ref{t_emd}. We see that the predicted $\kappa$ values from the SW2 potential in both directions are about four times as large as those from the SW1 potential. On the other hand, the anisotropy ratio, defined as $\kappa^{\rm zig}/\kappa^{\rm arm}$, is about four using both potentials. We note that our predicted $\kappa$ values using the SW2 potential parameterized by Xu \textit{et al.} \cite{Xu2015Direction} are more than two times larger than those obtained by Xu \textit{et al.} \cite{Xu2015Direction} using the EMD method. The reason for the difference is that the LAMMPS code \cite{plimpton1995jcp,lammps} used by them has a wrong implementation of the heat current for many-body potentials, as first pointed out in Ref. \onlinecite{fan2015prb} and then clearly demonstrated in Ref. \onlinecite{gill2015prb} for the Tersoff many-body potential. In Appendix \ref{appendix}, we explicitly demonstrate the incorrectness of the heat current computed with LAMMPS for the SW many-body potential.

\begin{table}[ht]
\caption{Thermal conductivity values (in units of W/mK) in the zigzag ($\kappa^{\rm zig}$) and armchair ($\kappa^{\rm arm}$) directions for SLBP at $300$ K from various methods. See text for the meaning of the acronyms EMD, HNEMD, NEMD, SW1, SW2, BTE, and DFT.}
\begin{center}
\begin{tabular}{ l l l l }
\hline
Reference & Method & $\kappa^{\rm arm}$ & $\kappa^{\rm zig}$\\
\hline
This work & EMD (SW1)   & $21 \pm 1$ & $92 \pm 4$ \\
This work & HNEMD (SW1) & $24 \pm 4$ & $97 \pm 4$ \\
This work & NEMD (SW1)  & $20 \pm 1$ & $92 \pm 2$ \\
Hong \textit{et al.}   \cite{Hong2015Thermal}  & NEMD (SW1) & $63.6 \pm 3.9$  & $110.7 \pm 1.75$  \\
Zhang \textit{et al.}  \cite{Zhang2016Thermal} & NEMD (SW1) & 9.89            & 42.55  \\
This work & EMD (SW2)   & $92 \pm 8$ & $360 \pm 30$ \\
This work & HNEMD (SW2) & $96 \pm 5$ & $361 \pm 2$ \\
Xu \textit{et al.}     \cite{Xu2015Direction}  & EMD (SW2)  & 33.0            &  152.7\\
Qin \textit{et al.}    \cite{qin2016prb}       & BTE (DFT)  & 4.59            & 15.33  \\
\hline
\end{tabular}
\end{center}
\label{t_emd}
\end{table}

\subsection{HNEMD results}

\begin{figure}[htb]
\begin{center}
\includegraphics[width=\columnwidth]{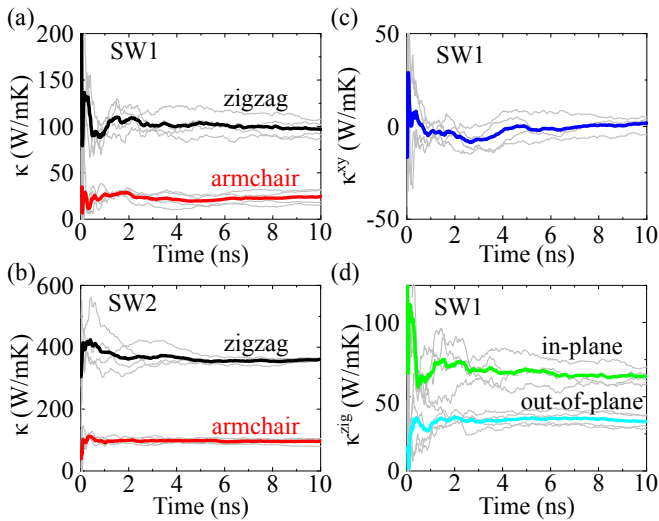}
\caption{Running average of the thermal conductivity  as a function of time from the HNEMD simulations. (a) $\kappa^{\rm zig}$ and $\kappa^{\rm arm}$ using the SW1 potential; (b) $\kappa^{\rm zig}$ and $\kappa^{\rm arm}$ using the SW2 potential; (c) $\kappa^{xy}$ using the SW1 potential; (d) the in-plane and  out-of-plane components of $\kappa^{\rm zig}$ using the SW1 potential. In each subplot, the thin lines are from four independent runs and the thick lines are the averages. }
\label{f_hnemd}
\end{center}
\end{figure}

The running averages $\overline{\kappa}(t)$ calculated using the HNEMD method are shown in Figs. \ref{f_hnemd}(a) and (b). Because the heat current (instead of the HCACF) is directly measured in this method, the variation between independent runs becomes smaller and smaller with increasing time. From the four independent values at $t=10$ ns, we obtain the $\kappa$ values and their error estimates for different transport directions and potentials. These values are also listed in Table \ref{t_emd}. It can be seen that the HNEMD and EMD results agree with each other very well. 

The HNEMD method can also be used to calculate the off-diagonal elements of the thermal conductivity tensor. For example, we can calculate $\kappa^{xy}$ as $\langle J^x(t)\rangle_{\rm ne}/TVF_{\rm e}^y$. We show the $\kappa^{xy}$ results for the SW1 potential in Fig. \ref{f_hnemd}(c). It can be seen that $\kappa^{xy}=0$, which means that the zigzag and armchair directions are the principal directions of the thermal conductivity tensor. The full thermal conductivity tensor in any coordinate system can thus be obtained from $\kappa^{\rm zig}$ and $\kappa^{\rm arm}$ using a coordinate transform. When the coordinate system is rotated counterclockwise by an angle of $\theta$ to a primed $x'y'$ coordinate system, the thermal conductivity tensor in the new primed coordinate system can be computed straightforwardly:
\begin{equation}
\kappa =
\left(
\begin{array}{cc}
\kappa^{\rm zig}\cos^2\theta + \kappa^{\rm arm}\sin^2\theta  &
\frac{1}{2}(\kappa^{\rm arm} - \kappa^{\rm zig} ) \sin2\theta \\
\frac{1}{2}(\kappa^{\rm arm} - \kappa^{\rm zig} ) \sin2\theta &
\kappa^{\rm zig}\sin^2\theta + \kappa^{\rm arm}\cos^2\theta \\
\end{array}
\right).
\end{equation}
In particular, when $\theta=\pi/4$, the off-diagonal element attains the maximum absolute value of $(\kappa^{\rm zig}-\kappa^{\rm arm})/2$.

It is sometimes useful to decompose the total thermal conductivity into some smaller contributions \cite{lv2016njp,matsubara2017jcp}. For 2D materials, one can decompose \cite{Fan2017Thermal} the heat current into an in-plane component and an out-of-plane component, corresponding to the in-plane phonons and the out-of-plane (flexural) phonons, respectively. The total thermal conductivity is then decomposed into an in-plane part and an out-of-plane part. From Fig. \ref{f_hnemd}(d), we see that the flexural phonons in SLBP contribute less than the in-plane phonons, which is opposite to the case of graphene \cite{Fan2017Thermal}.

\subsection{NEMD results}

\begin{table}[ht]
\caption{Thermal conductivity values (in units of W/mK) for systems with different lengths  $L$ (in units of nm) from the NEMD simulations using the SW1 potential. The labels ``periodic'' and ``fixed'' refer to the two simulation setups as shown in Fig. \ref{f_nemd}. }
\begin{center}
\begin{tabular}{ c c c c c c c c c}
\hline
$L$ & $\kappa^{\rm arm} (\rm periodic)$ & $\kappa^{\rm zig} (\rm periodic)$ & $\kappa^{\rm arm} (\rm fixed)$ & $\kappa^{\rm zig} (\rm fixed)$ \\
\hline
200 & $8.0  \pm 0.1$ & $38.2  \pm 0.7$  & $11.0  \pm 0.1$  & $56.7 \pm 0.6$ \\  
300 & $9.7  \pm 0.3$ & $48.4  \pm 0.6$  & $13.0  \pm 0.4$  & $64   \pm 1$ \\  
400 & $11.1 \pm 0.1$ & $55.4  \pm 0.9$  & $14.1  \pm 0.7$  & $69   \pm 2$ \\  
500 & $11.9 \pm 0.3$ & $60.1  \pm 0.4$  & $15.3  \pm 0.7$  & $73   \pm 1$ \\  
600 & $12.6 \pm 0.3$ & $62.7  \pm 0.9$  & $15.0  \pm 0.6$  & $76.5   \pm 0.8$ \\  
700 & $13.1 \pm 0.2$ & $66.1  \pm 0.7$  & $16.8  \pm 0.5$  & $78   \pm 1$ \\  
800 & $13.5 \pm 0.5$ & $67.9  \pm 0.8$  & $16.6  \pm 0.7$  & $80   \pm 2$ \\
\hline
\end{tabular}
\end{center}
\label{t_femd}
\end{table}
\begin{figure}[htb]
\begin{center}
\includegraphics[width=\columnwidth]{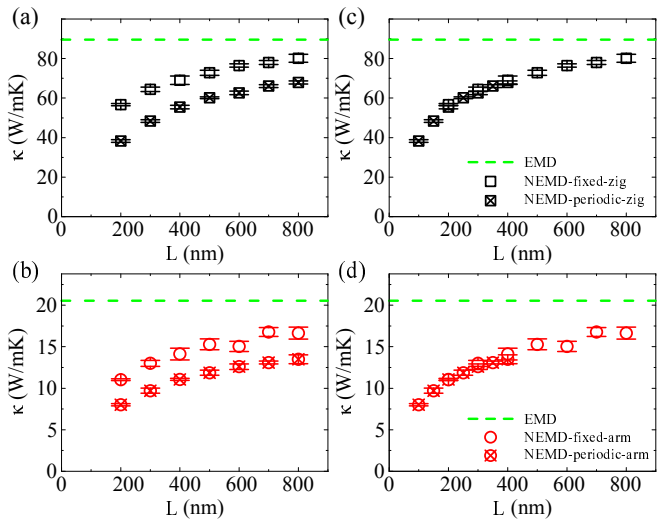}
\caption{Thermal conductivity values for SLBP at 300 K with different lengths $L$ from the NEMD simulations using the SW1 potential. The labels ``periodic'' and ``fixed'' refer to the two simulation setups as shown in Fig. \ref{f_nemd}. The system length $L$ in the periodic boundary setup is set to the simulation cell length in (a) and (b) and to half of the cell length in (c) and (d).}
\label{f_nofit}
\end{center}
\end{figure}

We calculated the thermal conductivities $\kappa(L)$ for seven system lengths from $200$ nm to $800$ nm and the results are summarized in Table \ref{t_femd} and visualized in Fig \ref{f_nofit}. We only consider the SW1 potential, which is the one adopted in previous works \cite{Hong2015Thermal,Zhang2016Thermal} using the NEMD method. We used two simulation setups shown schematically in Fig. \ref{f_nemd}. When we use $L$ as the system length for both setups, the $\kappa$ values from the periodic boundary setup are consistently smaller than those from the fixed boundary setup. When we change the system length in the periodic setup to $L/2$, the $\kappa$ values from both setups correlate with each other very well. This means that the system length should be identified as the source-sink distance rather than the simulation cell length. Therefore, it is less efficient to use the periodic boundary setup in the NEMD method. The periodic boundary step was also found \cite{azizi2017carbon} to be less efficient than the fixed boundary setup in obtaining converged Kapitza thermal resistance across graphene grain boundaries.

\begin{figure}[htb]
\begin{center}
\includegraphics[width=\columnwidth]{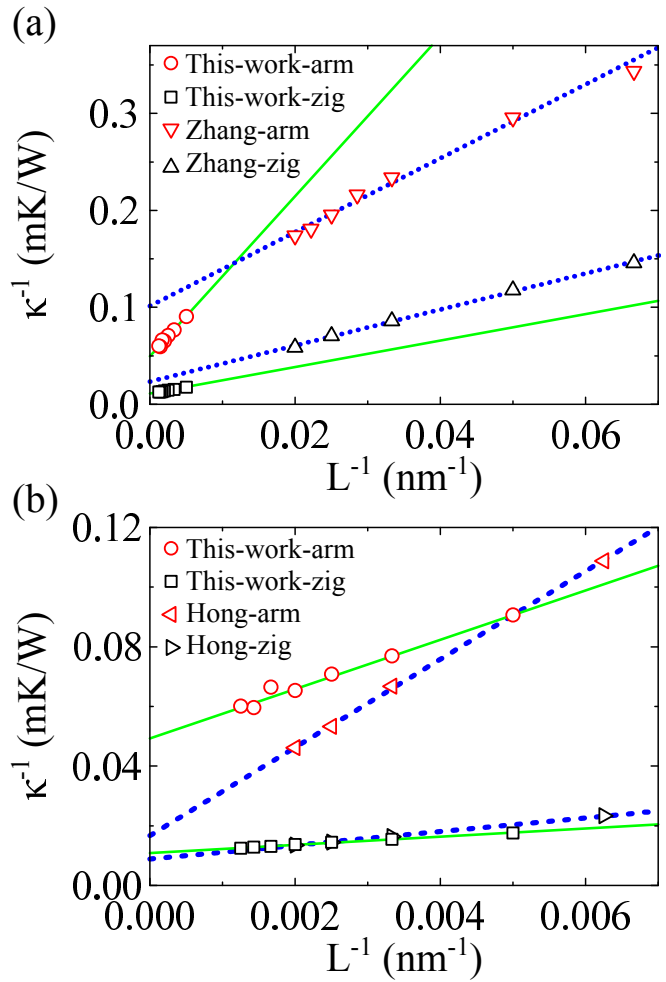}
\caption{Inverse of the thermal conductivity $1/\kappa(L)$ of SLBP at 300 K from NEMD simulations as a function of the inverse system length $1/L$. Our data (``This work'') are compared with those by Hong \textit{et al.} \cite{Hong2015Thermal} (``Hong'') and Zhang \textit{et al.} \cite{Zhang2016Thermal} (``Zhang''). Markers are NEMD data and lines are fits according to Eq. (\ref{e_inverse_k}). Because Zhang \textit{et al.} \cite{Zhang2016Thermal} used the periodic boundary setup, we have calculated the system length in their simulations as half of the simulation cell length.}
\label{f_comji}
\end{center}
\end{figure}

To compare the NEMD results with the EMD and HNEMD results, we need to extrapolate the NEMD values to the limit of infinite system length $\kappa(L=\infty)\equiv \kappa_0$. It was found \cite{dong2018prb} that when the system lengths are comparable and larger than the average phonon mean free path $\lambda$, the following linear relation \cite{schelling2002prb} between the inverse thermal conductivity and inverse length  holds well:
\begin{equation}
\frac{1}{\kappa(L)} = \frac{1}{\kappa_0} \left(1 + \frac{\lambda}{L}\right).
\label{e_inverse_k}
\end{equation}
Here, we use the NEMD data with the fixed setup where the system length is the simulation cell length $L$. Figure \ref{f_comji} shows the $1/\kappa(L)$ values against $1/L$, along with the fit according to Eq. (\ref{e_inverse_k}). The fitted values for $\kappa_0^{\rm zig}$ and $\kappa_0^{\rm arm}$ are listed in Table \ref{t_emd}. They are consistent with our EMD and HNEMD results, in line with the conclusion in Ref. \onlinecite{dong2018prb}. The fitted $\lambda$ values are about $170$ nm in the armchair direction and $130$ nm in the zigzag direction, which explains why the linear relation Eq. (\ref{e_inverse_k}) is valid for our NEMD data with $L \geq 200$ nm. 

\begin{figure}[htb]
\begin{center}
\includegraphics[width=\columnwidth]{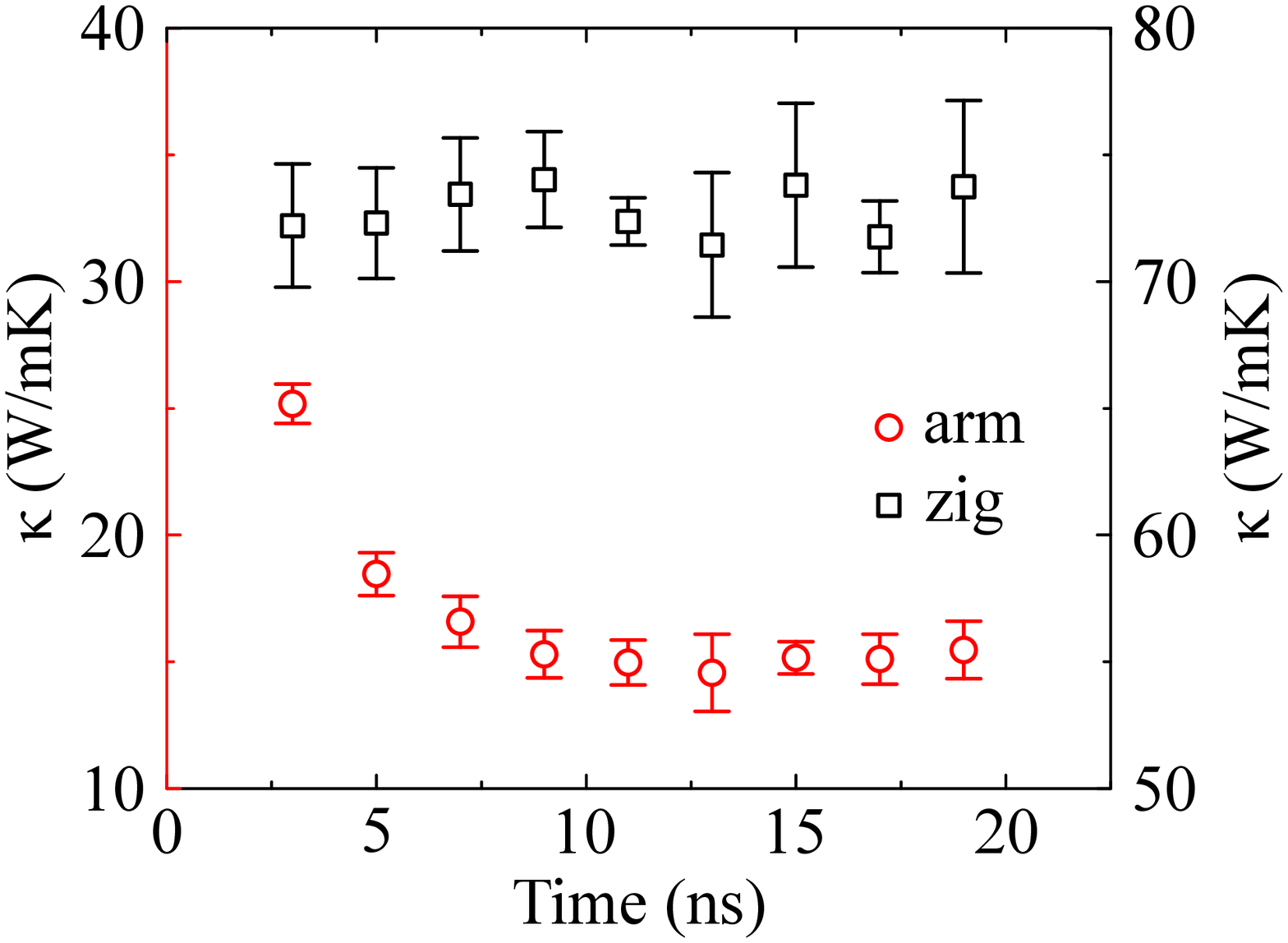}
\caption{Thermal conductivity in the armchair (left axis) and zigzag (right axis) directions in the 500-nm-long system as a function of the average of the time interval within which the thermal conductivity is evaluated. Each time interval spans $2$ ns. }
\label{f_nemd_time}
\end{center}
\end{figure}

We note that the predictions by Hong \textit{et al.} \cite{Hong2015Thermal} and Zhang \textit{et al.} \cite{Zhang2016Thermal} (listed in Table \ref{t_emd}) using the NEMD method and the SW1 potential differ significantly from ours. From Fig. \ref{f_comji}(a), we see that Zhang \textit{et al.} \cite{Zhang2016Thermal} used  systems much shorter than $\lambda$ to make the linear extrapolation, which is known \cite{dong2018prb,sellan2010prb} to be inappropriate. Hong \textit{et al.} \cite{Hong2015Thermal} considered systems up to $500$ nm, but their NEMD data for the armchair direction are not consistent with ours, as shown in Fig. \ref{f_comji}(b). To understand this discrepancy, we note that Hong \textit{et al.} \cite{Hong2015Thermal} has only used $4$ ns for the heat current generation stage in the NEMD simulation, which is not enough for long systems. To show this, we take the $500$-nm-long system as an example. Figure \ref{f_nemd_time} shows the thermal conductivities calculated within every two ns ($2-4$ ns, $4-6$ ns, etc). We can see that the thermal conductivity in the zigzag direction converges quickly but that in the armchair direction only converges at about $t=10$ ns. Hong \textit{et al.} \cite{Hong2015Thermal} did not mention the time interval used for measuring their thermal conductivity, but it is apparent that a simulation time of $4$ ns is not enough to bring the system into a steady state. Because the temperature gradient starts from zero when the heat current is generated, the temperature gradient in their simulation was underestimated and the thermal conductivity overestimated. 

\subsection{Performance evaluation of the MD based methods}

In the above, we have shown that with proper implementation and data analysis, consistent results with comparable error estimates can indeed be obtained using three rather different  MD based methods. However, they have very different computational costs. We measured the computational cost of each method in terms of the product of the number of atoms and the simulation time (sum of the equilibration time and production time). According to Sec. \ref{s_methods}, the computational costs (for one potential and two directions) in the EMD, HNEMD, and NEMD methods are about $1.2\times 10^8$, $2.7\times 10^6$, and $4.3\times 10^8$ atom $\cdot$ ns, respectively. Therefore, among the three methods, the NEMD method is the most inefficient and the HNEMD method is the most efficient. In particular, the HNEMD method is almost two orders of magnitude more efficient than the EMD method, consistent with the conclusion in Ref. \onlinecite{fan2018submitted}.

\subsection{Comparison with BTE results}

\begin{figure}[htb]
\begin{center}
\includegraphics[width=\columnwidth]{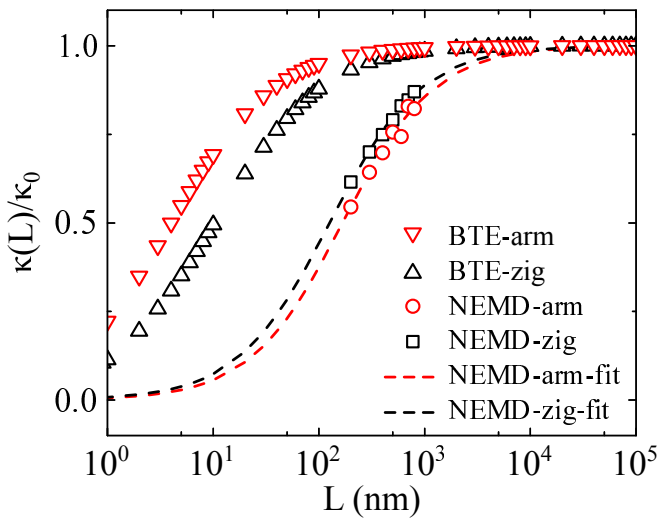}
\caption{The normalized thermal conductivity $\kappa(L)/\kappa_0$, where  $\kappa_0=\kappa(L=\infty)$, as a function of the system length $L$. Our NEMD results (NEMD data represented by squares and circles and fits using Eq. (\ref{e_inverse_k}) represented by dashed lines) are compared with the BTE results by Qin \textit{et al.} (triangles) \cite{qin2016prb}.}
\label{f_md_vs_bte}
\end{center}
\end{figure}

The SW1 potential by Jiang \cite{jiang2015nt} and the SW2 potential by Xu \textit{et al.} \cite{Xu2015Direction} have identical functional forms and both were parameterized based on their phonon structure data with the help of the same fitting code. However, as we can see from Table \ref{t_emd}, the predicted thermal conductivity values differ by about a factor of four. This difference suggests that the SW potential may not be able to reliably describe the interactions in SLBP. Actually, the early predictions \cite{zhu2014prb,Ong2014Strong,Qin2015Anisotropic,Jain2015Strongly} using the BTE based method also differ from each other by several times. Recently, Qin \textit{et al.} \cite{qin2016prb} found that long-range interactions in SLBP  caused by the resonant bonding play an important role in the phonon structure and transport properties. They showed that the thermal conductivity calculated using the BTE approach decreases with increasing cutoff distance and only converges up to about $7$ \AA, which is much larger than the cutoff distance (about $3$ \AA) in the SW potentials \cite{jiang2015nt,Xu2015Direction}. Their predicted thermal conductivity values (listed in Table \ref{t_emd}) are about an order of magnitude smaller than our MD predictions. From the length scaling of the normalized thermal conductivity $\kappa(L)/\kappa_0$ shown in Fig. \ref{f_md_vs_bte}, we see that the average phonon mean free paths from our NEMD simulations (with the SW1 potential) are about an order of magnitude larger than those from the BTE based calculations by Qin \textit{et al.} \cite{qin2016prb}. This comparison highlights the inadequacy of the short range SW potential in describing the phonon anharmonicity in SLBP. 

\section{Summary and conclusions}

In summary, we have computed the in-plane thermal conductivity of SLBP using three MD based methods, including the EMD method, the HNEMD method, and the NEMD method, and obtained consistent results. Among the three methods, we find that the HNEMD method is the most efficient and the NEMD method is the most inefficient. We also find that the system lengths in the NEMD method with the periodic boundary setup should be taken as half of the simulation cell lengths in order to make the calculated thermal conductivity values consistent with those obtained by using the NEMD method with the fixed boundary setup. Our main results are listed in Table \ref{t_emd}， where some previous data using MD and BTE calculations are presented for comparison. Previous MD results are erroneous due to various reasons: using an incorrect heat current formula as implemented in LAMMPS in the EMD method \cite{Xu2015Direction}, considering too short simulation times in the NEMD method \cite{Hong2015Thermal}, or considering too short system lengths in the NEMD method \cite{Zhang2016Thermal}. The thermal conductivity values and average phonon mean free paths from our MD simulations are about an order of magnitude larger than the most recent predictions obtained using the BTE approach considering long-range interactions in DFT calculations. This suggests that the short-range SW potential might be inadequate for describing the phonon anharmonicity in SLBP.

\begin{figure}[htb]
\begin{center}
\includegraphics[width=\columnwidth]{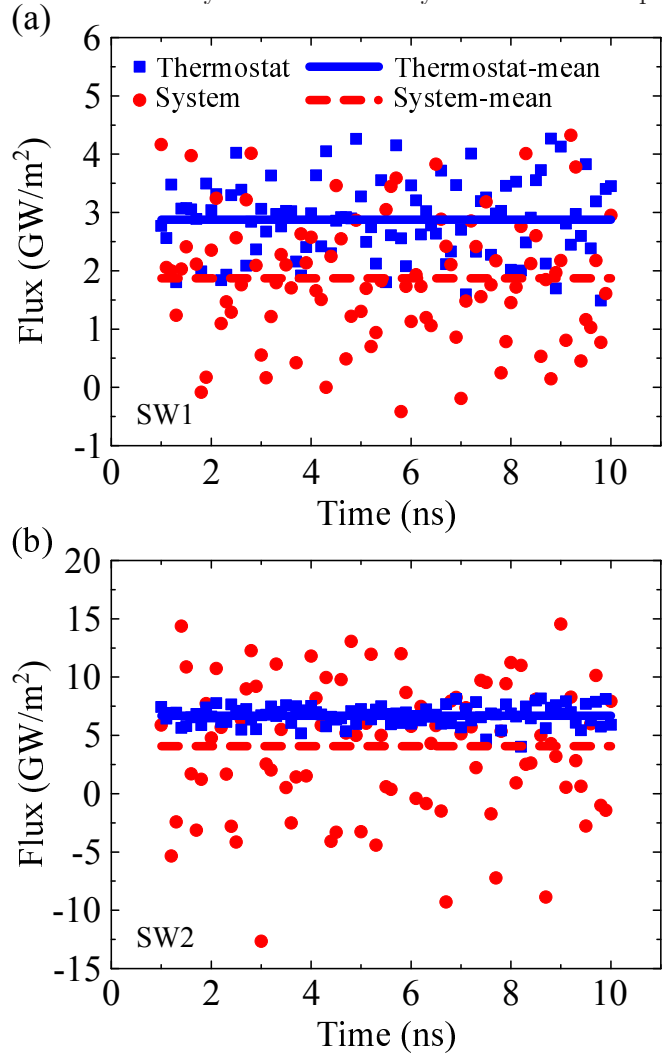}
\caption{Heat flux as a function of production time in the heat current generation stage of the NEMD simulations with the SW1 (a) and SW2 (b) potentials. The squares and solid lines represent the results computed from the energy exchange between the thermostats and the heat source and sink regions. The circles and dashed lines represent the results computed from the system (excluding the heat source and sink regions) using the heat current formula as documented in the LAMMPS manual \cite{lammps}.}
\label{f_heat_current}
\end{center}
\end{figure}

\begin{acknowledgments}
We thank Guangzhao Qin for reading the draft of the manuscript and for helpful comments.
This work was supported in part by the National Natural Science Foundation of China (Grant Nos. 11404033 and 11502217) and in part by the Academy of Finland QTF Centre of Excellence program (Project 312298). We acknowledge the computational resources provided by Aalto Science-IT project, Finland's IT Center for Science (CSC) and HPC of NWAFU.
\end{acknowledgments}

\appendix

\section{Demonstration of the error in the heat current computed with LAMMPS\label{appendix}}

As has been pointed out in Ref. \onlinecite{fan2015prb}, The heat current in LAMMPS \cite{plimpton1995jcp,lammps} is calculated from the virial stress tensor, which is only applicable to two-body potentials and leads to underestimated thermal conductivity for two-dimensional materials described by many-body potentials. This has been explicitly demonstrated by Gill-Comeau and Lewis \cite{gill2015prb} in terms of the Tersoff many-body potential. On the other hand, the heat current as implemented in the GPUMD code \cite{fan2017cpc,gpumd} is applicable to general many-body potentials, as has also been demonstrated in terms of energy conservation \cite{gill2015prb,Fan2017Thermal}. In this Appendix, we explicitly show that the heat current as implemented in the LAMMPS code \cite{plimpton1995jcp,lammps} is incorrect for the SW many-body potential.

To this end, we use LAMMPS \cite{plimpton1995jcp} to do NEMD simulations (in the fixed boundary setup) with a simulation cell of length $100$ nm and width $20$ nm, considering both the SW1 and the SW2 potentials. Here, the transport is along the armchair direction and the heat source and sink regions are maintained at 330 K and 270 K, respectively. The heat flux calculated from the thermostats and that from the particles in the system (excluding the source and sink regions) are compared in Fig. \ref{f_heat_current}.  According to energy conservation, the two heat fluxes should be the same when the system is in a steady state. The discrepancy between them as shown in Fig. \ref{f_heat_current} demonstrates that the LAMMPS implementation leads to an underestimation of the heat current and hence an underestimation of the thermal conductivity.

\bibliography{refs}

\end{document}